\documentclass[aip,graphicx]{revtex4-1}
\usepackage{graphicx}
\begin{document}


\title{Remote sensor response study in the regime
of the microwave radiation-induced magnetoresistance oscillations}


\author{Tianyu Ye}
\author{R. G. Mani}
\affiliation{Department of Physics and Astronomy, Georgia State University, Atlanta, Georgia 30303, USA}

\author{W. Wegscheider}
\affiliation{Laboratorium f\"ur Festk\"orperphysik, ETH Z\"urich, 8093 Z\"urich, Switzerland}

\date{\today}

\begin{abstract}
A concurrent remote sensing and magneto-transport study of the
microwave excited two dimensional electron system (2DES) at liquid
Helium temperatures has been carried out using a carbon detector
to remotely sense the microwave activity of the 2D electron system
in the GaAs/AlGaAs heterostructure during conventional
magneto-transport measurements. Various correlations are observed
and reported between the oscillatory magnetotransport and the
remotely sensed reflection. In addition, the oscillatory remotely
sensed signal is shown to exhibit a power law type variation in
its amplitude, similar to the radiation-induced magnetoresistance
oscillations.
\end{abstract}


\maketitle

The ultra-high-mobility two dimensional electron system (2DES) in
GaAs/AlGaAs heterostructures  exhibits giant oscillatory
magnetoresistive response at liquid helium temperature in the form
of Microwave-induced Zero-resistance States
(MiZrS)\cite{Maninature2002} and Microwave-induced
Magnetoresistance Oscillations
(MIMOs)\cite{Maninature2002,ZudovPRLDissipationless2003} under
photo-excitation. Such
response is potentially useful for electromagnetic wave
characterization in the technologically useful microwave and
terahertz bands.\cite{ManiAPL2008} Thus, photo-excited
magnetotransport has been extensively studied by experiment over
the past
decade.\cite{ManiPRBVI2004,ManiPRLPhaseshift2004,ManiEP2DS152004,KovalevSolidSCommNod2004,SimovicPRBDensity2005,ManiPRBTilteB2005,WiedmannPRBInterference2008,DennisKonoPRLConductanceOsc2009,ManiPRBPhaseStudy2009,ManiPRBAmplitude2010,ArunaPRBeHeating2011,ManiPRBPolarization2011,ManinatureComm2012,ManiPRBterahertz2013,TYe2013}
As well, many theories have been developed to explain the
MIMO's\cite{DurstPRLDisplacement2003,
AndreevPRLZeroDC2003,RyzhiiJPCMNonlinear2003,
KoulakovPRBNonpara2003,LeiPRLBalanceF2003, DmitrievPRBMIMO2005,
LeiPRBAbsorption+heating2005, InarreaPRLeJump2005,InarreaAPL2006,
ChepelianskiiEPJB2007,InarreaAPL2008_2,InarreaAPL2008,InarreaAPL2009,FinklerHalperinPRB2009,
ChepelianskiiPRBedgetrans2009, InarreaNanotech,
InarreaPRBPower2010,
MikhailovPRBponderomotive2011,Inarrea2011,Inarrea2012,Inarrea2013}
such as, for example, the displacement model that invokes
radiation-assisted indirect inter-Landau-level scattering by
phonons and
impurities,\cite{DurstPRLDisplacement2003,RyzhiiJPCMNonlinear2003,LeiPRLBalanceF2003}
the non-parabolicity model which considers non-parabolicity
effects in an ac-driven system,\cite{KoulakovPRBNonpara2003} the
inelastic model which explores the effects of a radiation-induced
steady state non-equilibrium
distribution,\cite{DmitrievPRBMIMO2005} and the radiation driven
electron orbit model which follows the periodic motion of the
electron orbit centers under
irradiation.\cite{InarreaPRLeJump2005} The oscillatory minima of
the MIMO's are thought to transform into the experimentally
observed zero-resistance states through either a current
instability,\cite{AndreevPRLZeroDC2003,FinklerHalperinPRB2009} or
from the accumulation/depletion of carriers at the
contacts.\cite{MikhailovPRBponderomotive2011}

Here, we use a remote sensing method\cite{TYe2013} to further
characterize MIMOs and study the evolution of the remotely sensed
reflection signal and transport response with microwave frequency
and microwave power, and compare the frequency and power
dependence of the remote sensing signal with the magnetoresistive
response of the 2DES. Our results also show a non-linear response
vs. the microwave power in the oscillatory remotely sensed signal,
similar to the observations of the non-linear oscillatory response
observed in MIMOs.\cite{ManiPRBAmplitude2010}

\begin{figure}[t]
\centering
\includegraphics[width= 85mm]{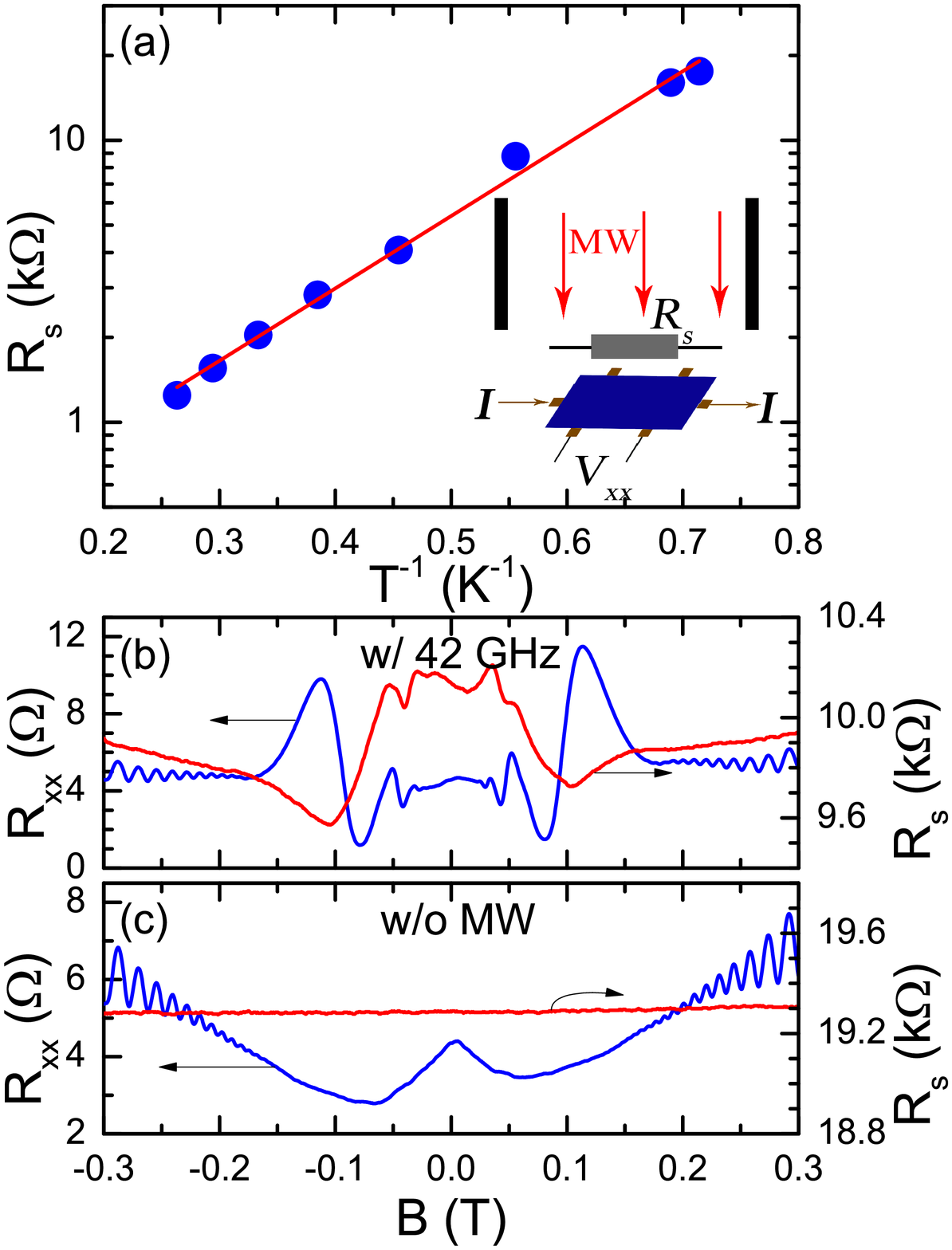}
\caption{(Color online) (a) The inset shows a sketch of our
experiment set-up. This panel shows the response curve of the
carbon sensor, $R_{s}$. The red line shows that the sensor's
resistance increases rapidly with the reduction of the
temperature. (b) and (c) exhibit sample's diagonal resistance
$R_{xx}$ (left ordinate) and sensor resistance $R_s$ (right
ordinate) as functions of magnetic field, B, with (w/) and without
(w/o) 42 GHz microwave excitation, respectively.}
\end{figure}

Experiments were carried out on high mobility GaAs/AlGaAs
heterostructure Hall bar with gold-germanium contacts at liquid
Helium temperatures. The specimen was mounted horizontally at the
end of a sample probe and inserted into variable temperature
insert inside the bore of superconducting solenoid. A base
temperature of approximately 1.5 K was realized by pumping on the
liquid helium within the variable temperature insert. Microwaves
were sent by a launcher from the top of a circular microwave
waveguide to the sample (Fig. 1(a) inset). A carbon remote sensor
was placed above- and in close proximity to- the sample. This
sensor exhibits a strong temperature coefficient, which is
attributed to activated transport, in its response at liquid
Helium temperatures, see Fig. 1 (a). After a brief illumination by
a red Light-emitting diode to realize the high mobility condition, standard four
terminal low frequency lock-in techniques were adopted to measure
the sample signal and sensor signal concurrently.

\begin{figure}[t]
\centering
\includegraphics[width=85 mm]{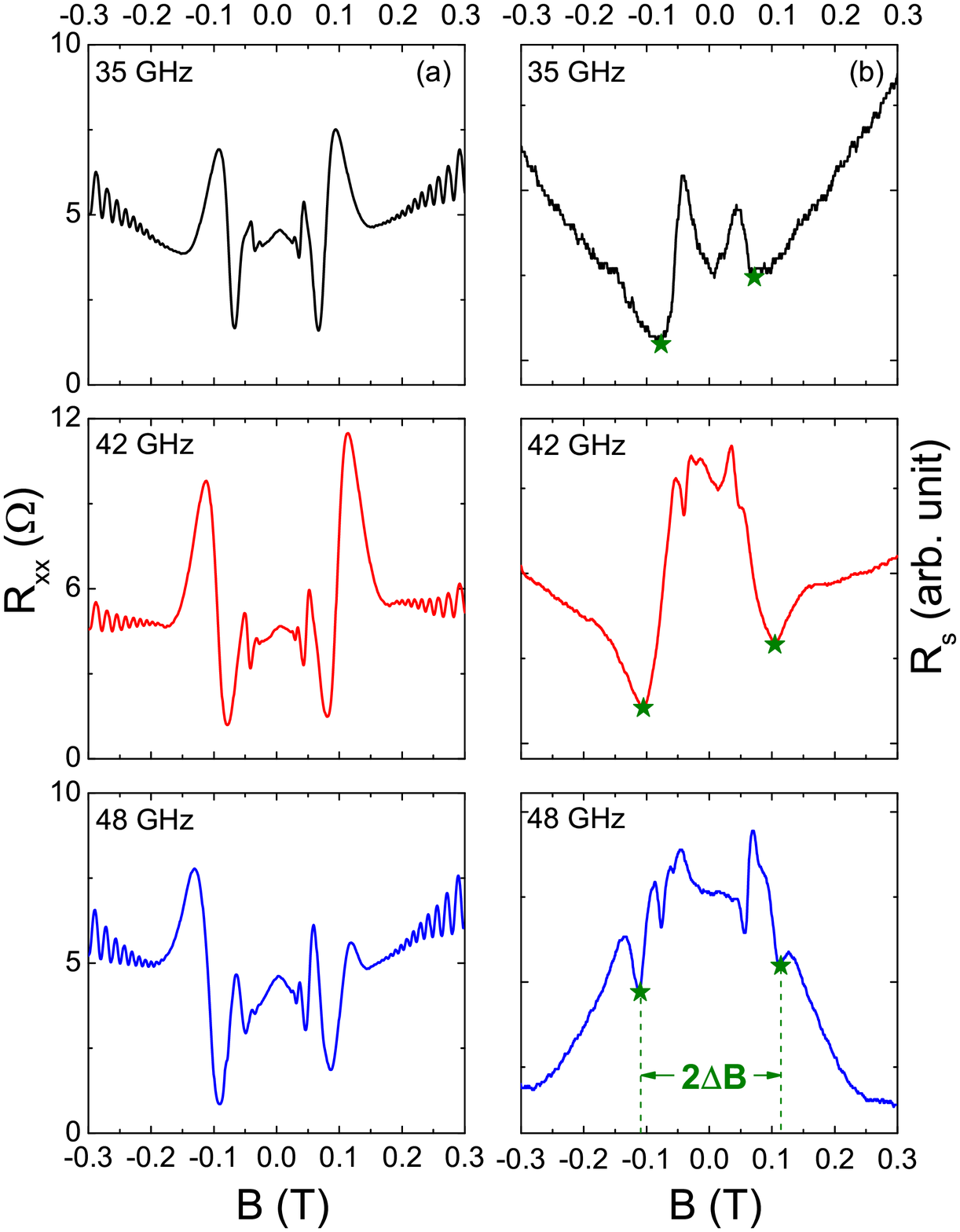}
\caption{(Color online) Column (a) shows the diagonal resistance,
$R_{xx}$, vs. the magnetic field, $B$, for various microwave
frequencies. Column (b) shows the corresponding sensor resistance,
$R_{s}$, vs. $B$. From top to bottom, the microwave frequencies
are 35 GHz, 42 GHz, and 48 GHz. Stars in column (b) mark the last
major valley in the $R_s$ traces. Bottom panel of column (b)
illustrates the determination of $2\Delta B$.}
\end{figure}

The carbon sensor was found to be extremely effective in detecting
small changes in the reflection associated with the MIMOs. As
shown in Fig. 1 (b), with 42 GHz microwave illumination, the
diagonal resistance $R_{xx}$ exhibits strong MIMOs within the
magnetic field span $-0.15 \le B\le +0.15$ Tesla, and the sensor
resistance $R_s$ also reveals oscillatory response within the same
regime. Without microwave excitation, both $R_{xx}$ and $R_s$ do
not exhibit any oscillatory features,  see Fig. 1(c), over the
$-0.15 \le B\le +0.15$ Tesla $B$-span. Thus, $R_s$ oscillations
are correlated with 2D electron transport induced by microwave
excitation. As a consequence, the $R_s$ signal is adopted here to
remotely sense the characteristics of MIMOs. It appears worth
pointing out that with- or without- microwaves, the zero-field
value of the magnetoresistance trace ($R_{xx}$)in 2DES is
approximately the same. On the other hand, the remote sensor
signal $R_s$ shifts by approximately 9 k$\Omega$ between the
microwave on- and off- conditions. Such a large shift in $R_s$ is
attributed to microwave heating, mostly due to the incident
radiation. Thus, shifts in $R_{s}$ do not serve to characterize
transport in the 2DES; it is only the oscillatory features on the
$R_s$ traces that reflect the detector response to the
microwave-induced transport in the 2DES.

Fig. 2 exhibits $R_{xx}$ and $R_s$ responses to different
microwave frequencies. Column (a) shows the $R_{xx}$ of the 2DES
as the microwave frequency is increased from 35 GHz to 48 GHz.
Note that MIMOs span a wider magnetic field regime as the
microwave frequency, $f$, increases because a larger Landau level
splitting is required to obtain energy commensurability as the
photon energy $hf$ increases. Column (b) shows the concurrently
measured sensor resistance $R_s$. Here, the oscillatory $R_s$ span
a wider magnetic field as the microwave frequency increases,
similar to MIMOs, see Fig. 2(a). At the same time, as the
microwave frequency increases, more oscillatory structure appears
in the $R_s$, similar to MIMOs. At all the frequencies, there are
discernable boundaries (marked by stars in Fig. 2 (b)), marking
the $B$-span for oscillatory $R_s$. Further, a step-like response
is apparent in the vicinity of the starred locations in Fig. 2(b).
Such step-like remote sensor response is similar to the electron
absorption reported by I\~narrea \textit{et
al.}\cite{InarreaNanotech}. This feature suggests that the remote
sensing signal is correlated to the response of the 2D electron
system to microwave excitation.

\begin{figure}[t]
\centering
\includegraphics[width= 85mm]{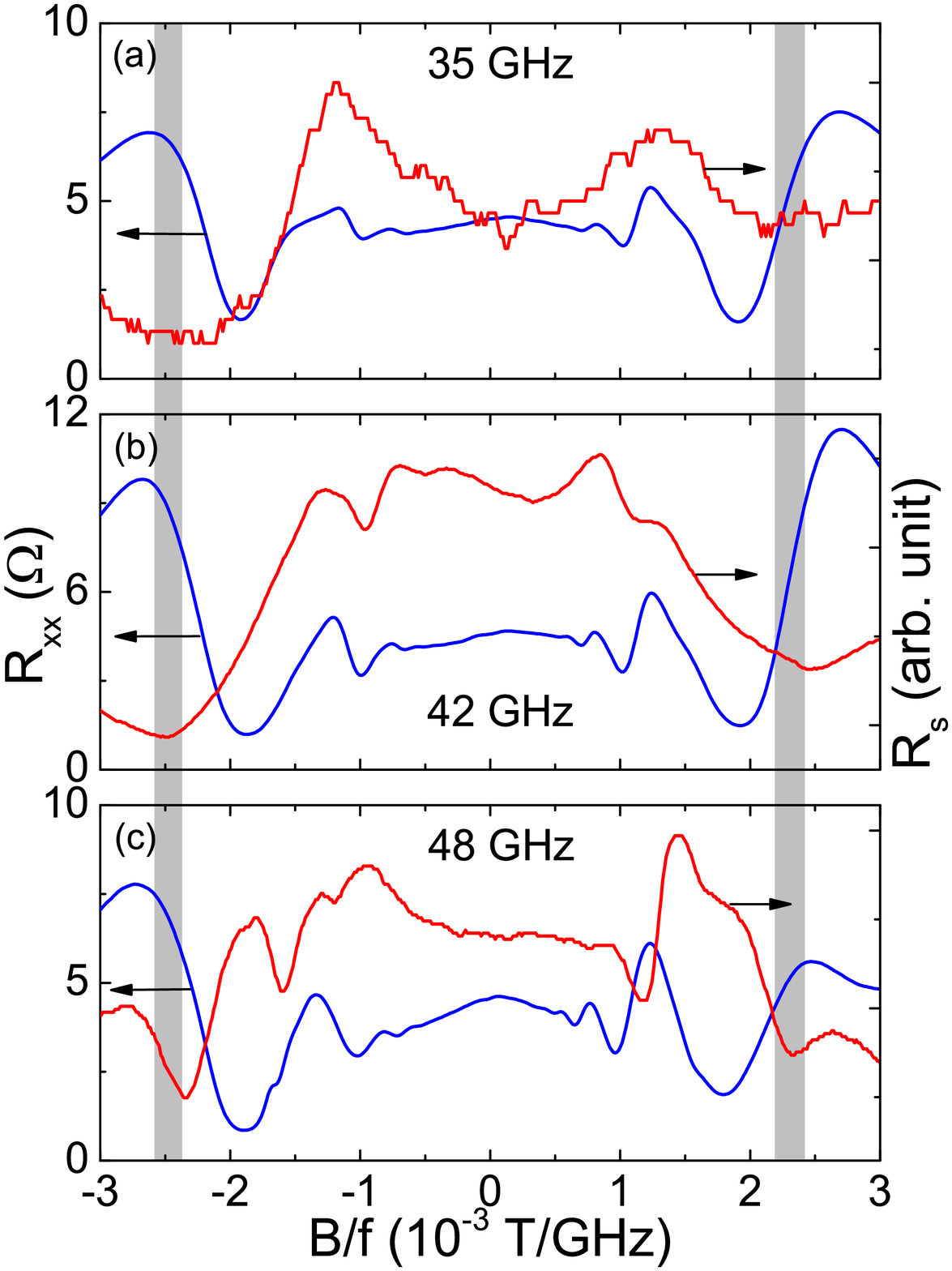}
\caption{(Color online) Panels (a) to (d) plot $R_{xx}$ (left
scale) and $R_s$ (right scale) vs. $B/f$ for various microwave
frequencies. Shadowed vertical bands mark the minima on $R_s$ traces.}
\end{figure}

Figs. 3 (a) - (c) shows $R_{xx}$ and $R_s$ traces as functions of
$B/f$ at three microwave frequencies. As the frequency increases,
more MIMOs reveal themselves in the $R_{xx}$ trace. For instance,
at 35 GHz, only three MIMOs appear for each direction of the
magnetic field. At 48 GHz, at least four oscillations are
discernable for each direction of the magnetic field. Note that,
in this $B/f$ plot, the MIMO's do not shift their positions on the
abscissa as microwave frequency change. It is amazing that similar
to the MIMOs, the turning points of $R_s$ on the abscissa also do
not shift with the microwave frequency; they are fixed within the
band $2.3 \times 10^{-3} \le B/f \le 2.5 \times 10^{-3}$ T/GHz,
slightly above cyclotron resonance. For $|B/f| < 2.3\times
10^{-3}$ T/GHz regime, the microwave energy, $hf$, satisfies, $hf
\ge \hbar \omega_{c}$, allowing for radiation-induced
inter-Landau-level transitions. Hence, the
oscillatory remote sensor signal appears when radiation-induced
inter-Landau-level transitions are allowed in the 2DES.

\begin{figure}[t]
\centering
\includegraphics[width= 85mm]{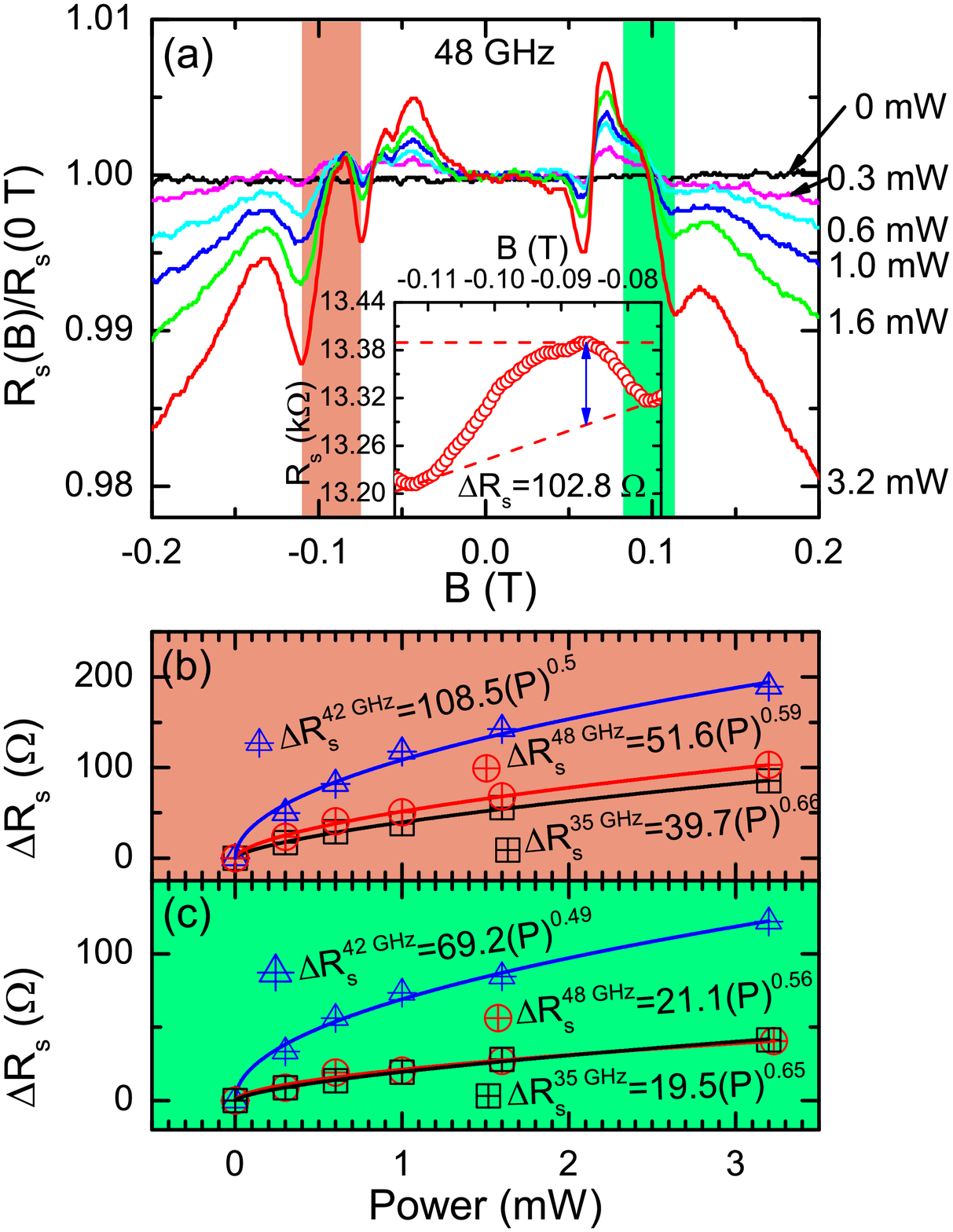}
\caption{(Color online) (a) This panel shows the power dependence
of $R_s$ vs. $B$ under 48 GHz microwave excitation. All $R_s$
curves are normalized to their own zero magnetic field values. The
inset of (a) illustrates the determination of the amplitude
$\Delta R_s$ of the $R_{s}$ oscillation at 3.2 mW power in the
brown area. The arrowed blue line indicates the $\Delta R_s$
associated with this oscillation. Panels (b) and (c) show plots of
$\Delta R_s$ as functions of microwave power. Panel (b)
corresponds to the oscillations in the brown shaded area in panel
(a) and panel (c) corresponds to the green shaded area in panel
(a). Different symbols indicate different frequencies: squares for
35 GHz, triangles for 42 GHz, and circles for 48 GHz. Power law
functions are adopted to fit the symbols to solid lines. The fit
results are also given in panels (b) and (c).}
\end{figure}

Using the above results, we have shown a correlation between MIMOs
and the remotely sensed reflection signal. Below, we examine the
evolution of oscillatory features in the reflection signal with
the microwave power, see Fig. 4. Since the $R_{s}$ signal is
also sensitive to the heating produced by the incident microwaves
as mentioned previously, we plot zero-magnetic-field-normalized
$R_s$ values, i.e., $R_{s}(B)/R_{s}(B=0)$ in Fig. 4(a) for the
sake of presentation. In Fig. 4 (a), as the microwave power
increase from 0 to 3.2 mW, the amplitude of oscillatory features
in $R_{s}(B)/R_{s}(B=0)$ becomes larger, but their relative
positions on the abscissa are fixed. We evaluated the amplitudes
(without normalization) of the oscillatory features from the
traces with different microwave powers. As indicated in Fig. 4(a),
we measured the vertical height or amplitude from peak to
the base line of one oscillation and plot these values as a
function of microwave power in Fig. 4(c) and (d), respectively,
for oscillatory features at negative and positive magnetic fields.
The power dependence of the amplitude of oscillatory $R_s$ for all
the frequencies could be fit with a power law function $\Delta
R_s(f) =AP^{\alpha}$, where $A$ and $\alpha$ are fit parameters
that vary with the microwave frequency, see Fig. 4 (b) and (c). We
found that the $\alpha$ value for every frequency was close to
$1/2$. Moreover, the $\alpha$ value for oscillatory features at
postivie and negative magnetic fields are almost the same: 0.66
(-$B$) and 0.65 (+$B$) for 35 GHZ; 0.5 (-$B$) and 0.49 (+$B$) for 42 GHz;
and 0.59 (-$B$) and 0.56 (+$B$) for 48 GHz.

Both the frequency and power dependence of the remotely sensed
reflection signal suggest that the remotely sensed reflection
signal $R_s$ could serve to monitor in real-time the microwave
induced transport in the GaAs/AlGaAs 2DES. In addition, the
microwave power dependence measurements indicate a non-linear
microwave power dependence of the remotely sensed signal $\Delta
R_s$, which agrees with the experimental report of the power
dependence of $R_{xx}$ in the 2DES\cite{ManiPRBAmplitude2010}.
Indeed, both $R_{xx}$ and $\Delta R_{s}$ could be fit with the
same power law function with exponent approximately equal to
$1/2$. Such agreement further supports the reliability of remotely
sensing the microwave-induced transport in the 2DES.

Finally, we examine our results utilizing the radiation driven
electron orbit model\cite{InarreaPRLeJump2005}, because the
simulations made with this model
\cite{InarreaAPL2009,InarreaNanotech,InarreaPRBPower2010} appear
consistent with our results. In this model, radiation forces the
electron orbit center to move back and forth in the direction of
the radiation electric field at the frequency of radiation, and
the $R_{xx}$ oscillations reflect the periodic motion of the
electron orbit center. According to this model, the amplitude of
diagonal resistance oscillations $R_{xx}\propto E_0$, the
amplitude of the microwave electric field. Since the radiation
power $P=$$1\over 2$$c_{GaAs}\epsilon_{GaAs}\epsilon_0E_0^2$,
where $c_{GaAs}$ and $\epsilon_{GaAs}$ are the speed of light and
dielectric constant in GaAs, and $\epsilon_0$ is the dielectric
constant in vacuum, this theory asserts that the amplitude of
oscillatory $R_{xx}\propto \sqrt{P}$, implying an exponent of
$1/2$ in plots such as Fig. 4(b) and Fig. 4(c), close to the
experimentally observed value. Further, the microwave absorption
model-simulation of this theory indicates a sharp change of the
microwave absorption in 2DES in the vicinity of cyclotron
resonance, which is consistent with the observations reported here
in connection with figure 3.

In summary, we utilized a carbon sensor to remotely sense the
photo-excited transport properties of 2D electrons in the regime
of MIMOs. By changing the microwave frequency and power applied to
the specimen, we deduced correlations between the observed
magnetotransport in the 2DES and the remotely sensed reflection
signal. We have also observed that oscillatory features in the
remotely sensed reflection signal exhibit a non-linear amplitude
variation with the microwave power, similar to the power-law type
variation reported for the oscillatory diagonal resistance
associated with MIMOs.

Basic research and helium recovery at Georgia State University is
supported by the U.S. Department of Energy, Office of Basic Energy
Sciences, Material Sciences and Engineering Division under
DE-SC0001762. Additional support is provided by the ARO under
W911NF-07-01-015.

\pagebreak
\section*{Figure Captions}
Figure 1: (Color online)(a) This panel shows the response curve of
carbon sensor. The red line shows that the sensor's resistance
increases rapidly with the reduction of the temperature. The inset
shows a sketch of our experiment set-up. (b) and (c) exhibit
sample's diagonal resistance $R_{xx}$ (left ordinate) and sensor
resistance $R_s$ (right ordinate) as functions of magnetic field,
B, with and without (w/o) 42 GHz microwave excitation.

Figure 2: (Color online) Column (a) shows the diagonal resistance,
$R_{xx}$, vs. the magnetic field, $B$, for various microwave
frequencies. Column (b) shows the corresponding sensor resistance,
$R_{s}$, vs. $B$. From top to bottom, the microwave frequencies
are 35 GHz, 42 GHz, and 48 GHz. Stars in column (b) mark the last
major valley in the $R_s$ traces. Bottom panel of column (b)
illustrates the determination of $2\Delta B$.

Figure 3: (Color online) Panels (a) to (d) plot $R_{xx}$ (left
scale) and $R_s$ (right scale) vs. $B/f$ for various microwave
frequencies. Shadowed vertical bands mark the minima on $R_s$ traces.

Figure 4: (Color online) (a) This panel shows the power dependence
of $R_s$ vs. $B$ under 48 GHz microwave excitation. All $R_s$
curves are normalized to their own zero magnetic field values. The
inset of (a) illustrates the determination of the amplitude
$\Delta R_s$ of the $R_{s}$ oscillation at 3.2 mW power in the
brown area. The arrowed blue line indicates the $\Delta R_s$
associated with this oscillation. Panels (b) and (c) show plots of
$\Delta R_s$ as functions of microwave power. Panel (b)
corresponds to the oscillations in the brown shaded area in panel
(a) and panel (c) corresponds to the green shaded area in panel
(a). Different symbols indicate different frequencies: squares for
35 GHz, triangles for 42 GHz, and circles for 48 GHz. Power law
functions are adopted to fit the symbols to solid lines. The fit
results are also given in panels (b) and (c). \pagebreak

\begin{figure}[t]
\centering
\includegraphics[width=85 mm]{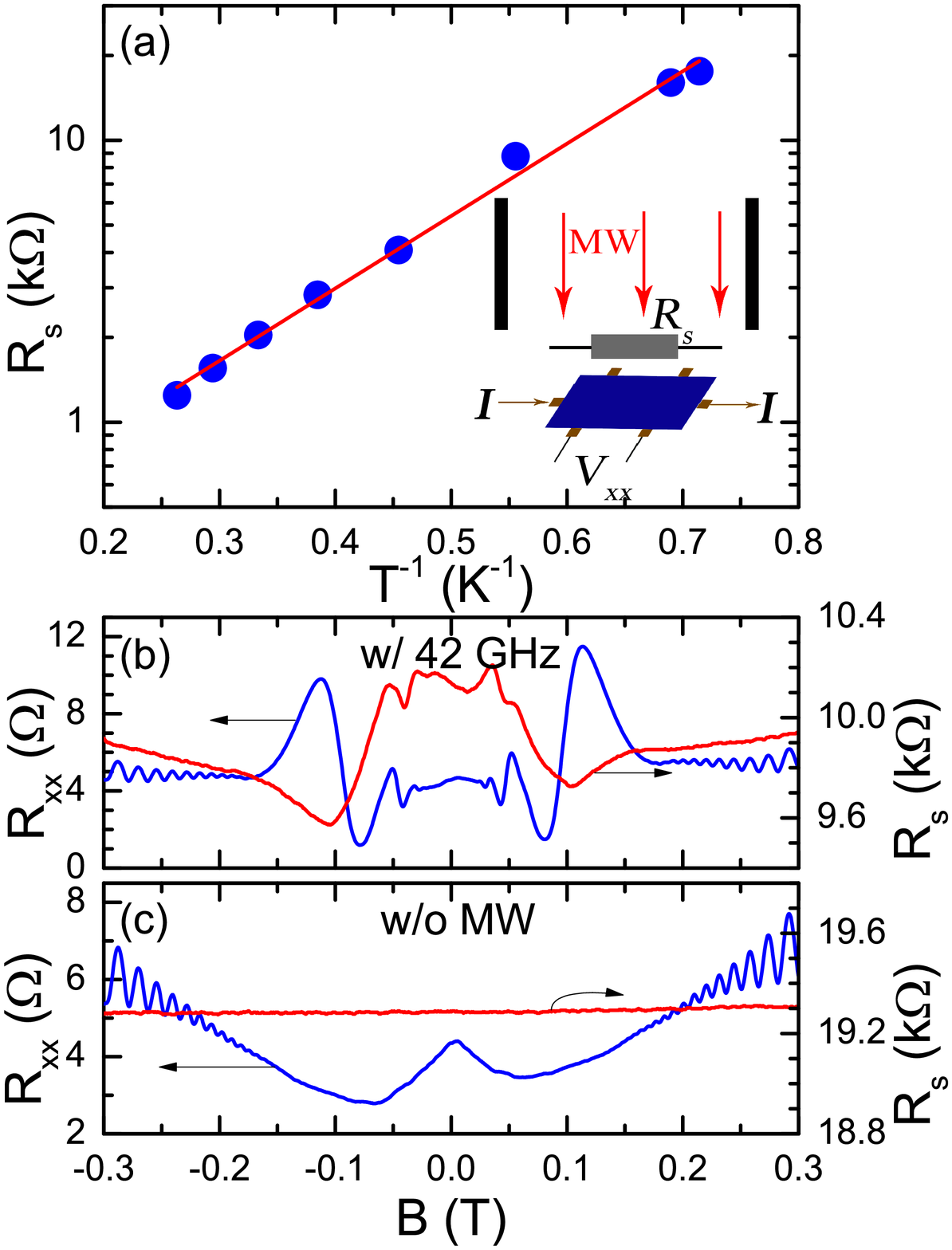}
\begin{center}
Figure 1
\end{center}
\end{figure}

\begin{figure}[t]
\centering
\includegraphics[width=85 mm]{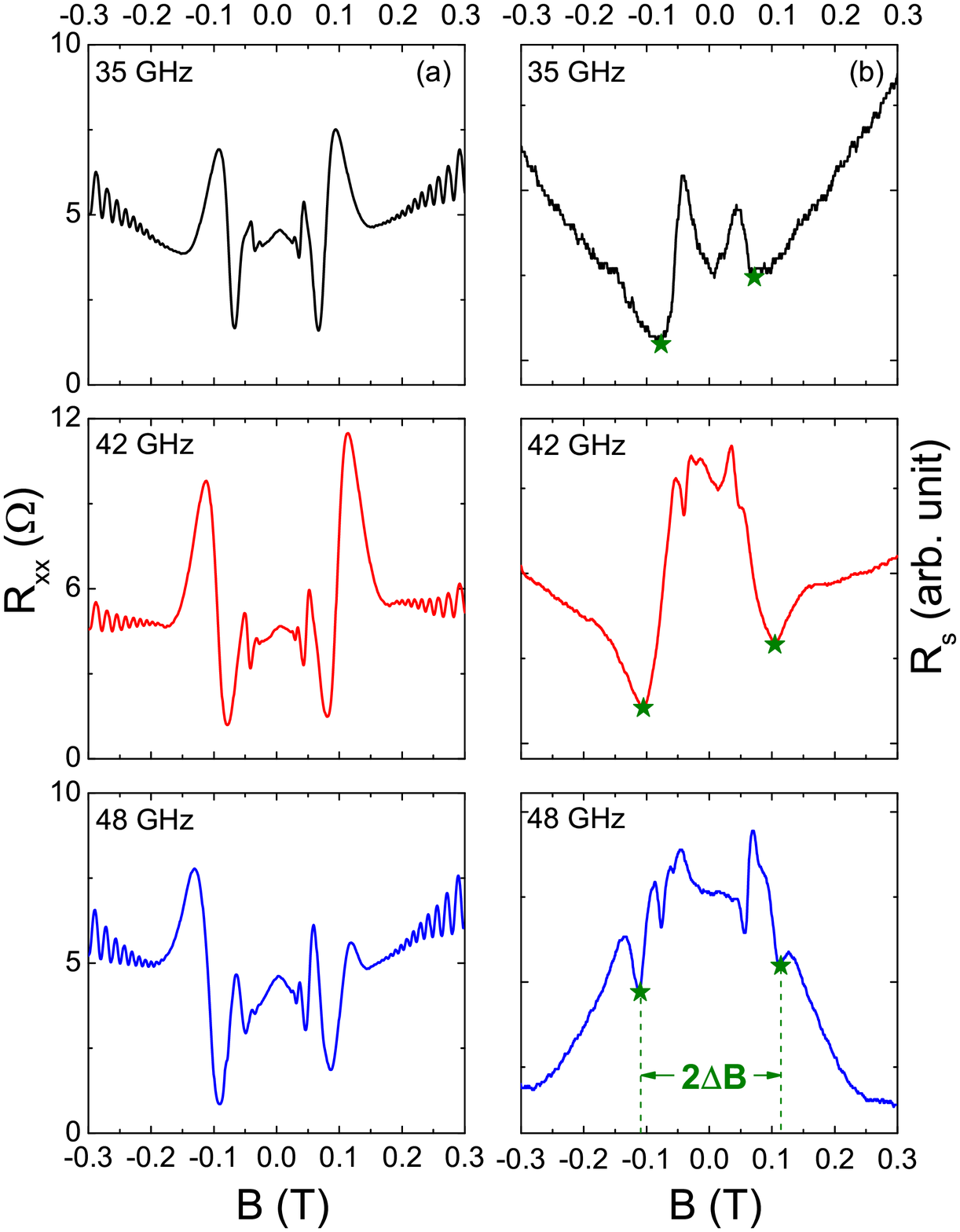}
\begin{center}
Figure 2
\end{center}
\end{figure}

\begin{figure}[t]
\centering
\includegraphics[width=85 mm]{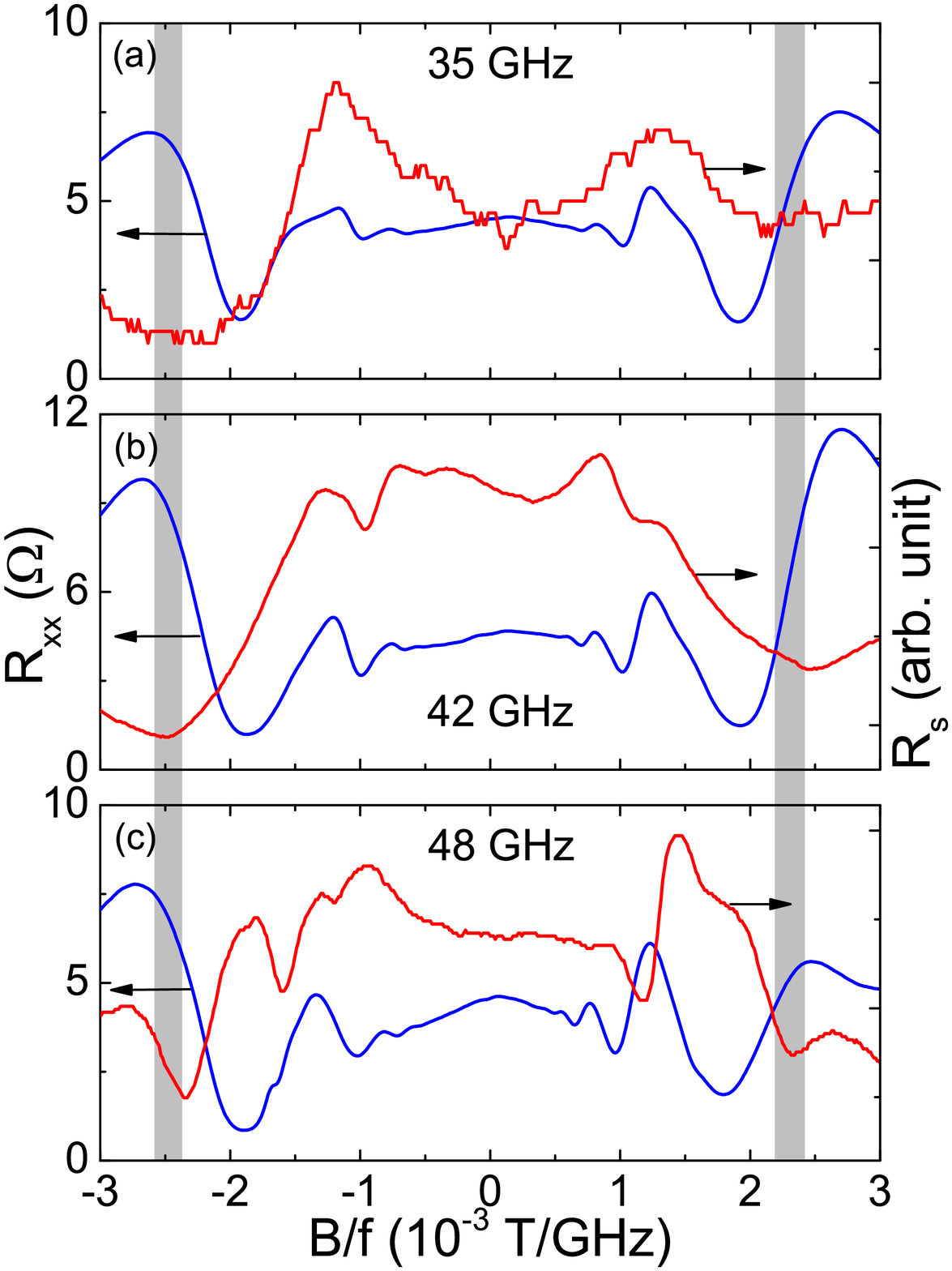}
\begin{center}
Figure 3
\end{center}
\end{figure}

\begin{figure}[t]
\centering
\includegraphics[width=85 mm]{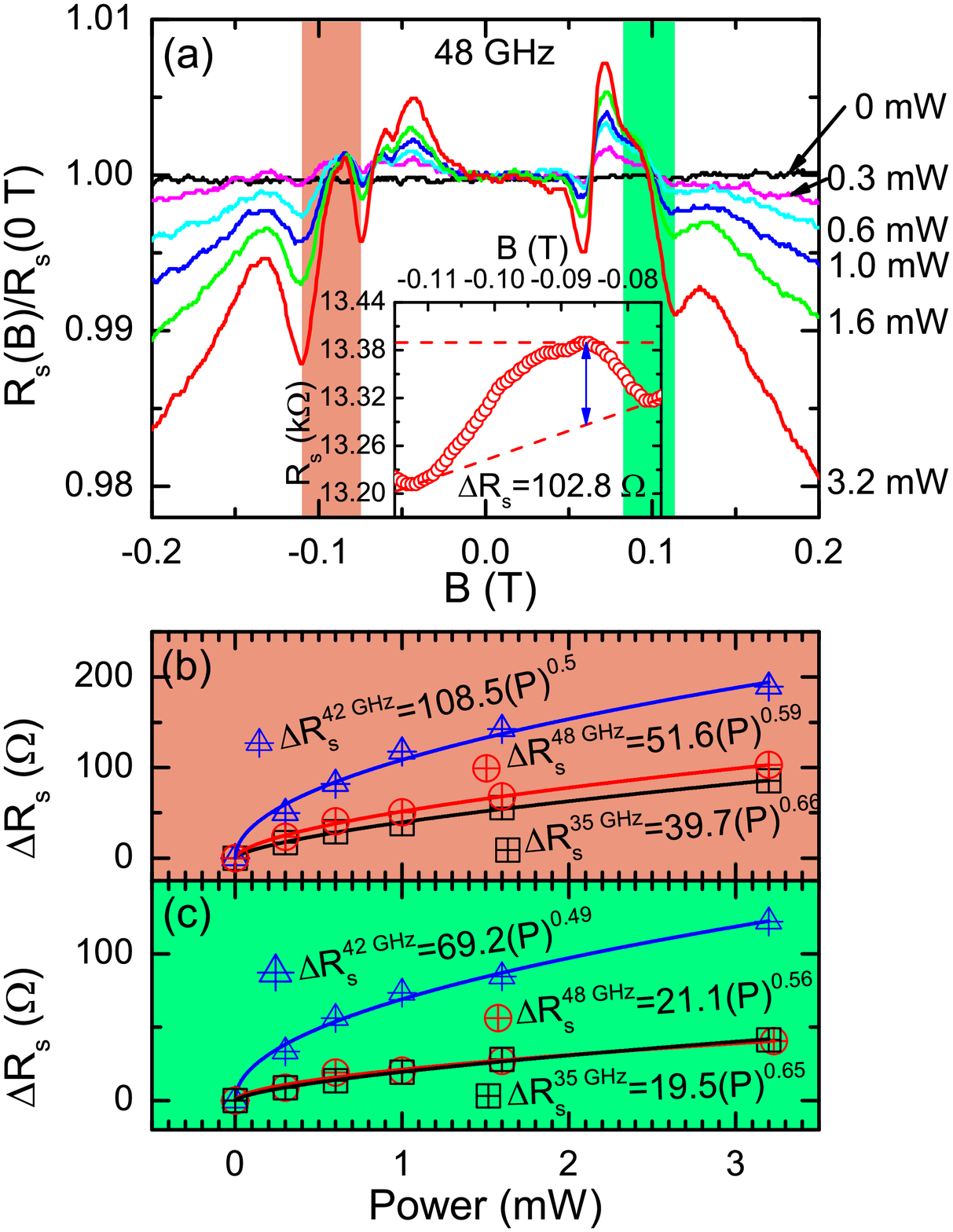}
\begin{center}
Figure 4
\end{center}
\end{figure}

\end{document}